\documentclass{JINST}

\makeatletter
\let\ifpdf\relax
\makeatother

\usepackage{amsmath,graphicx,float}
\usepackage[normalem]{ulem}
\usepackage{color}
\usepackage{url}
\usepackage{cleveref}
\usepackage[caption=false]{subfig}
\usepackage{booktabs}
\usepackage{siunitx}
\usepackage{lineno}
\usepackage{textcomp}
\title{Industrial Deposition of Wavelength--Shifting Films for Liquid Argon Photon Detection Systems}

\author{Babak Azmoun$^a$, Aleksey Bolotnikov$^b$, Francesca Capocasa$^b$, Milind Diwan$^a$, Yimin Hu$^c$, Jay Hyun Jo$^a$\thanks{Corresponding author.}, William Lenz$^a$, Yichen Li$^a$, Abdul Rumaiz$^d$, Vyara Tsvetkova$^e$, and Matteo Vicenzi$^a$\\
\llap{$^a$}Physics Department, Brookhaven National Laboratory, Upton, NY 11973, USA\\
\llap{$^b$}Instrumentation Department, Brookhaven National Laboratory, Upton, NY 11973, USA\\
\llap{$^c$}LaserFiberOptics LLC, Clarksville, MD 21029, USA\\
\llap{$^d$}National Synchrotron Light Source II, Brookhaven National Laboratory, Upton, NY 11973, USA\\
\llap{$^e$}Department of Physics and Astronomy, Wellesley College, Wellesley, MA 02481, USA\\
E-mail: \email{jjo@bnl.gov}
}

\abstract{
The Deep Underground Neutrino Experiment (DUNE) Phase-II Far Detector is considering an approximately 2000\,m$^2$ photon detection system to achieve a target mean light yield of 180\,PE/MeV. Meeting this requirement demands scalable, cost-effective, and high-quality wavelength-shifter (WLS) coatings capable of converting 127\,nm liquid-argon scintillation light into visible photons with controlled and reproducible optical performance. We report on the successful realization of an industrial physical vapor deposition (PVD) process for \textit{p}-terphenyl (pTP) coatings, adapted from vacuum deposition techniques developed for OLED display manufacturing, to produce uniform WLS layers on large-area inorganic substrates, a task traditionally challenged by adhesion and uniformity issues at organic--inorganic interfaces. Surface characterization by profilometry and spectroscopic measurements demonstrates edge-region thickness variation below 10\% and emission spectra consistent with high-quality pTP reference samples. The industrial process demonstrates reproducibility, scalability, and significantly reduced production time compared to laboratory-based methods, while maintaining optical characteristics consistent with established pTP reference samples. These results establish a viable pathway for mass production of high-performance pTP coatings for DUNE FD3 and future neutrino experiments, from a coating manufacturing and process standpoint. Detector-level performance validation, including quantitative VUV conversion efficiency measurements at 127\,nm, is identified as future work.}


\begin{document}

\section{Introduction}\label{sec:intro}

The Deep Underground Neutrino Experiment (DUNE)~\cite{DUNE:2020lwj} is a next-generation long-baseline neutrino oscillation experiment designed to address fundamental questions in particle physics, including leptonic charge conjugation parity (CP) violation, the neutrino mass ordering, and proton decay. Its photon detection system (PDS) plays a critical role in detecting scintillation light from liquid-argon time-projection chambers (LArTPCs), enabling precise event timing, improved calorimetric energy reconstruction, and enhanced sensitivity to low-energy physics channels~\cite{PhysRevD.111.032007, f3t7-5vmd}.

\begin{figure}[t]  
  \subfloat[X-Arapuca baseline geometry\label{fig:xarapuca}]{
    \includegraphics[width=0.45\textwidth]{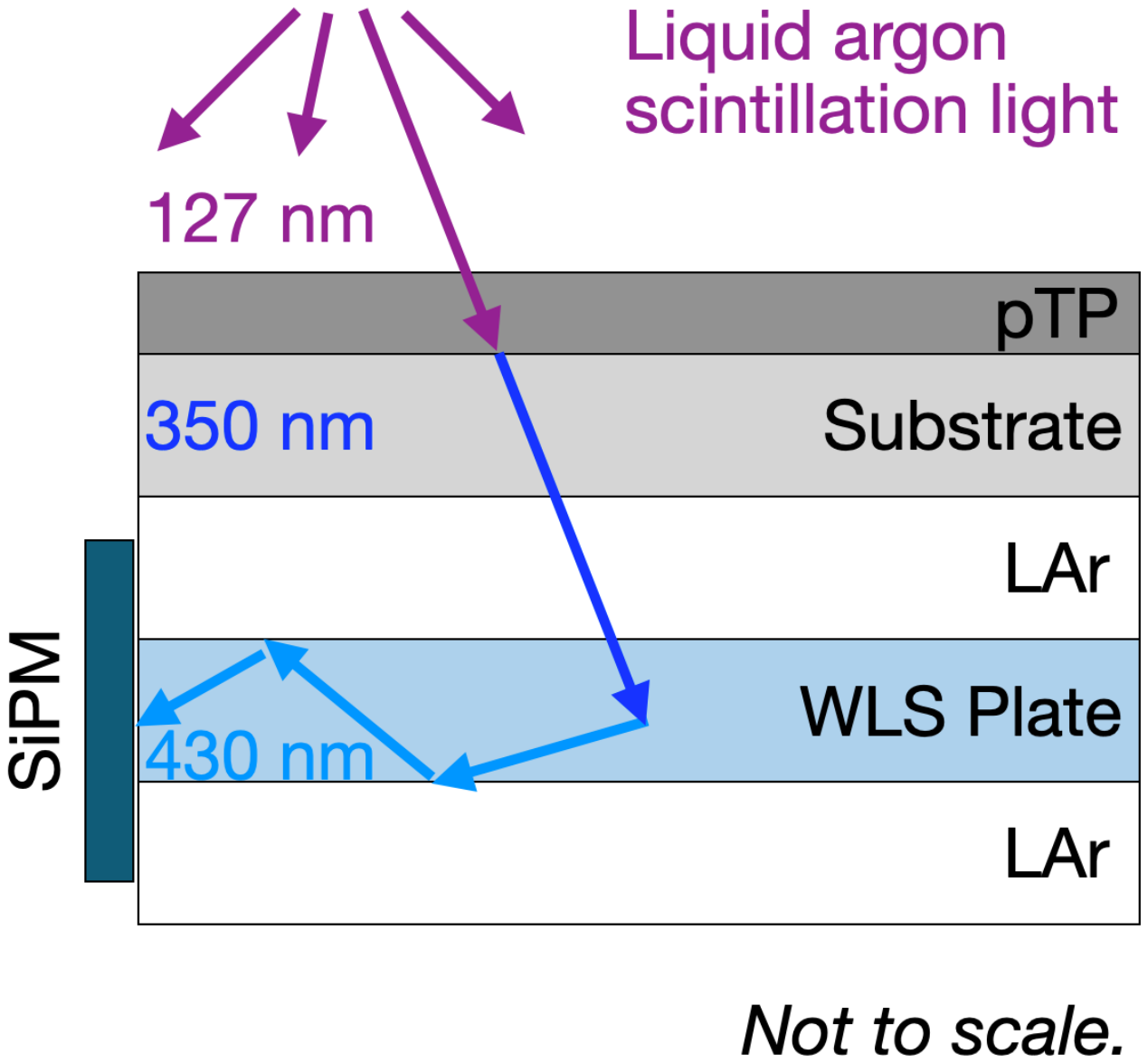}
  }\hfill
  \subfloat[pTP absorption and emission spectra\label{fig:ptp_spectrum}]{
    \includegraphics[width=0.5\textwidth]{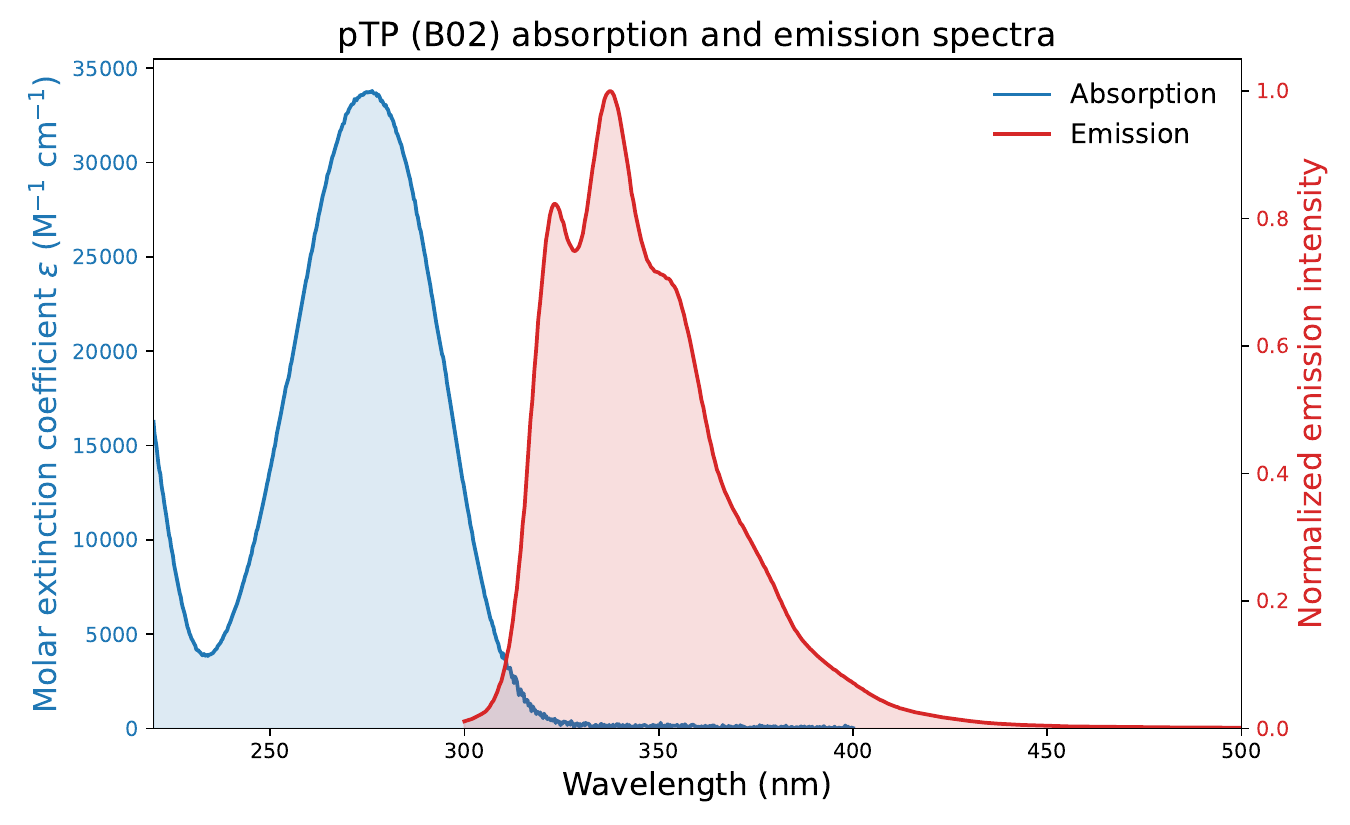}
  }
  \caption{
  (a) Working principle of an X-Arapuca cell, based on Ref.~\cite{FALCONE2021164648}. (b) Absorption and emission spectra of pTP, generated from data obtained from PhotochemCAD~\cite{PhotochemCAD}. The blue curve shows the excitation light peaking at 266~nm, and the red curve shows the emission spectrum of pTP.
  }
\end{figure}

In the DUNE Phase-I Far Detectors 1 and 2 (FD1/2)~\cite{DUNE:2020lwj, DUNE:VD_tdr}, the baseline light-collection technology is the X-Arapuca light-trap design~\cite{xarapuca}. This concept combines a primary wavelength-shifting (WLS) layer, an optical trapping mechanism with a secondary WLS layer, and silicon photomultiplier (SiPM) readout to achieve high photon-collection efficiency. Figure~\ref{fig:xarapuca} illustrates the working principle of the X-Arapuca cell. In this configuration, the primary WLS is deposited directly onto a substrate facing the LAr volume, where it converts the 127\,nm VUV scintillation photons into near-UV photons suitable for subsequent wavelength shift, light trapping, and detection.

For DUNE Phase-II Far Detector~\cite{dune:phase2}, the PDS design remains under active optimization, and several alternative light-trap geometries are being investigated. Preliminary optical simulations and prototype studies indicate that comparable photon-collection performance may be achievable even without a dichroic interface, motivating parallel development of simplified single-layer WLS coatings. The present work focuses on realizing such a coating through industrial-scale deposition of \textit{p}-terphenyl (pTP) films on large-area substrates, evaluating their optical and mechanical properties from a manufacturing standpoint. The scope of this work is the demonstration of a scalable industrial deposition process and characterization of the resulting films' optical and geometric properties; detector-level performance validation, including quantitative VUV conversion efficiency at 127\,nm, is identified as future work.

pTP is selected as the primary WLS material for its high quantum efficiency, fast re-emission time, and emission spectrum peaking in the near-UV ($\sim$340--360\,nm), well matched to the subsequent optical elements~\cite{PhotochemCAD} (see Fig.~\ref{fig:ptp_spectrum}). The quality of the pTP layer directly impacts the photon-detection efficiency through its conversion efficiency, emission-spectrum stability, and optical coupling to the trapping system.

Coating an organic wavelength-shifting material such as pTP onto inorganic substrates (e.g., borosilicate glass, quartz, or sapphire) presents intrinsic challenges. Differences in surface energy, thermal expansion coefficient, and chemical bonding properties often lead to poor adhesion, microcracking, or non-uniform film morphology. These effects are particularly pronounced when scaling to large-area substrates, where even slight variations in surface condition can result in substantial differences in film thickness and optical performance. Addressing these interface-related challenges is therefore essential for ensuring reproducibility and long-term stability of the coatings in large cryogenic detectors, and motivates detailed surface characterization of adhesion, morphology, and uniformity.

A variety of wavelength-shifting (WLS) coating techniques have been developed for large-area photodetector applications. Vacuum thermal evaporation provides precise control over film thickness and excellent optical quality, making it widely used for small- and medium-scale coatings; however, the throughput is traditionally limited for very large substrates~\cite{Gehman2011TPB, Broerman2017DEAP}. Spin coating can produce smooth and uniform layers on moderately sized substrates, although thickness uniformity and edge control tend to degrade as the coating area increases~\cite{Yang2020SpinCoating}. Doctor-blade coating offers a scalable approach suitable for large surfaces but involves direct mechanical contact, which may introduce surface defects or non-uniform regions if not precisely controlled~\cite{Berni2004DoctorBlade}. Finally, solvent-based or spray-deposition techniques are simple and low-cost but typically require careful optimization of solvent choice, viscosity, and evaporation rate to achieve good adhesion and uniformity~\cite{Mavrokoridis2009ArDMWLS, Howard2018WLSPlates}. Each of these methods has been used successfully in laboratory-scale experiments, but industrial-scale implementation demands a process that can deliver reproducibility, high optical quality, and scalability without sacrificing coating precision.

For the Phase-II Far Detector-scale deployment, the pTP coating must meet stringent requirements. First, it should have a thickness of order 1\,\textmu m (with an acceptable range of roughly 1--2\,\textmu m) to ensure efficient absorption of 127\,nm light without unnecessary self-absorption or excess material, consistent with previous optimization studies of organic wavelength-shifting films~\cite{Gehman2011TPB, Broerman2017DEAP}. Second, the coating must achieve spatial uniformity at the $\lesssim 10\%$ level across each substrate, so that coating-induced light yield variations remain subdominant in the overall PDS response. Here, PDS response can be defined with minimum and average of the detector's light yield~\cite{marinho2025apex}. Third, the films must exhibit robust adhesion and mechanical stability to withstand handling, cryogenic cycling, and long-term immersion in LAr. In addition, the chemical and optical purity of the pTP layer must be sufficiently high to preserve the characteristic emission spectrum and avoid quenching or spectral distortions from impurities or degradation products. Scaling production to the $\mathcal{O}$(2000\,m$^2$) required for the Phase-II Far Detector motivates a transition from manual laboratory setups to industrialized processes that deliver high throughput, reproducibility, and in-line quality control.

The following sections describe the industrial coating process and characterization methods (Sec.~\ref{sec:methods}) and performance validation of pTP films on multiple substrate types (Sec.~\ref{sec:results}).

\section{Methods}\label{sec:methods}

The coating procedure developed in this work was established through collaboration between Brookhaven National Laboratory (BNL) and LaserFiberOptics LLC, adapting industrial physical vapor deposition (PVD) processes commonly used in the optical and display industries for the production of high-uniformity organic thin films. The goal was to demonstrate that such an industrial approach can be directly applied to the deposition of pTP wavelength-shifting layers for large-area photon detection systems in liquid-argon detectors. Compared to traditional laboratory-based techniques, this process offers improved reproducibility, precise thickness control, and the throughput necessary for scaling to square-meter--level production. The following sections describe the deposition setup, characterization methods, and substrate studies carried out to establish the process parameters and verify coating performance. These methods together established a reproducible industrial coating workflow with full traceability from process parameters to optical performance.

\subsection{Vacuum thermal deposition}\label{subsec:vacuum}

PVD encompasses a class of thin-film techniques wherein material is deposited onto a substrate in a low-pressure environment. Reducing the ambient pressure to the millitorr range (or lower) increases the mean free path of vaporized species well beyond chamber dimensions, enabling largely ballistic transport from the source to the substrate and minimizing gas-phase scattering. This facilitates dense, uniform coatings and precise control of film thickness and composition.

In thermal evaporation, a source material is heated---via resistive heating, electron-beam bombardment, or related methods---until it vaporizes and subsequently condenses on cooler surfaces. Film properties depend on deposition parameters such as base pressure, substrate temperature, deposition rate, and source--substrate geometry. In the absence of a carrier gas, vacuum-deposited films often exhibit high purity and low defect density relative to solution-processed coatings. Widespread adoption in microelectronics and optics reflects advantages in scalability, batch-to-batch reproducibility, and compatibility with metals, dielectrics, and organic compounds.

\subsection{Deposition procedure for pTP coatings}\label{subsec:procedure}

\paragraph{Pre-deposition substrate preparation}
Following the conventional standard protocols of optical substrate preparation, the substrates are cleaned in ultrasonic baths using solvents followed by deionized water rinses. These steps were found to be essential for achieving high-yield, defect-minimized pTP films.

\paragraph{Plasma surface treatment}

\begin{figure}[t]
  \centering
  \includegraphics[width=0.95\textwidth]{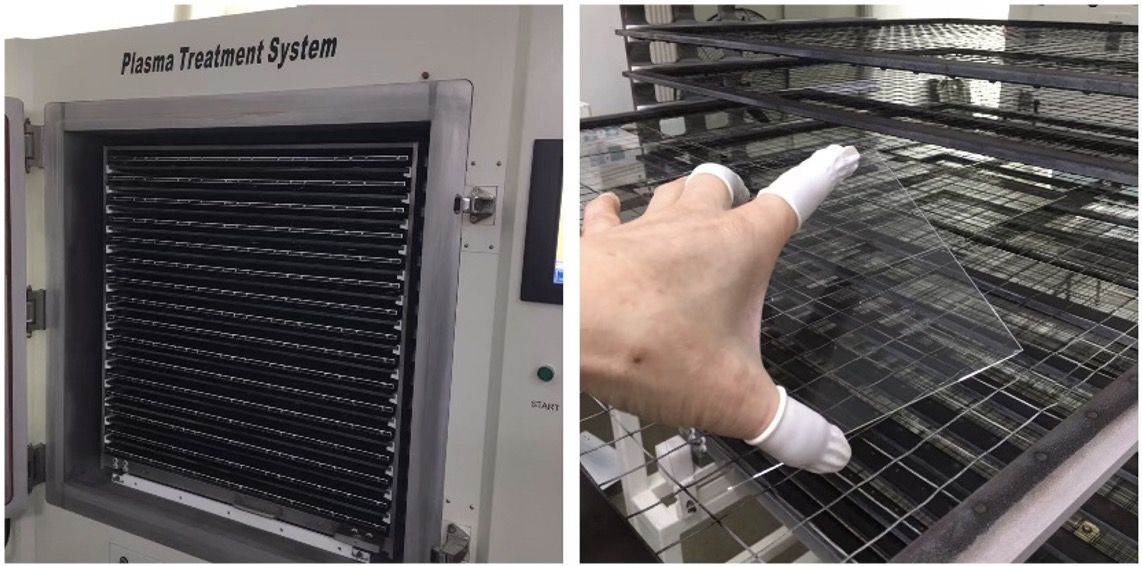}
  \caption{Plasma-treatment system and substrate placement. Mesh shadowing during treatment produced observable thickness modulation in early trials.}
  \label{fig:plasma_treatment}
\end{figure}

Plasma surface treatment was employed to enhance the adhesion and surface cleanliness of the substrates prior to pTP deposition. In this process, the substrate surfaces were exposed to a low-pressure reactive plasma that removes residual organic contaminants and modifies the surface chemistry, thereby improving bonding with the subsequently deposited pTP film. The plasma was generated from a mixture of N$_2$, O$_2$, and CF$_4$ gases in a ratio of 2:10:1 and applied for approximately 15\,minutes (Fig.~\ref{fig:plasma_treatment}). This gas mixture produces a chemically active plasma containing oxygen- and nitrogen-based species that promote surface activation, while CF$_4$ contributes mild fluorination and surface cleaning effects commonly used in optical and semiconductor processing. The substrates were placed on metal meshes inside the plasma chamber to ensure uniform exposure.

Areas partially shielded by the supporting mesh exhibited reduced plasma exposure, resulting in lower pTP adhesion and thinner coating in those regions. To avoid such grid-pattern artifacts in the final coatings, all subsequent samples were positioned so that the active deposition surface was fully exposed to the plasma. This configuration provided consistent surface activation and significantly improved coating uniformity across the entire substrate area.

\paragraph{In situ ion bombardment and deposition parameters}

The coater is equipped with an argon ion gun. Substrates mounted on a rotating dome were exposed to Ar ion flux for $\sim$10\,minutes prior to deposition. This treatment improved film nucleation and mitigated isolated regions of poor wetting attributable to trace organics. The ion flux also modestly elevated the substrate temperature.

\begin{figure}[t]
  \centering
  \includegraphics[width=0.55\textwidth]{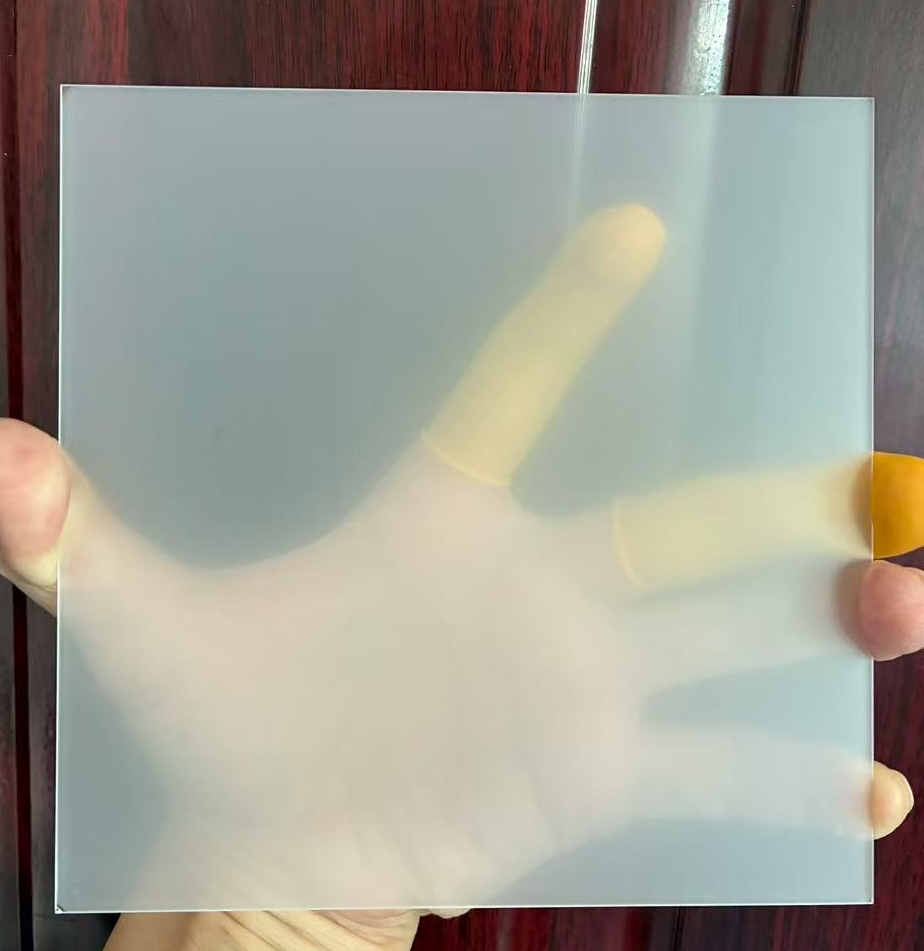}
  \caption{Uniform pTP film on a 143.75\,mm\,$\times$\,143.75\,mm B33 substrate.}
  \label{fig:deposition}
\end{figure}

Deposition was performed at a working pressure of $1.6\times10^{-4}$\,Torr (turbo-molecular pumping). The pTP source boat temperature was maintained at $\sim$195\,$^\circ$C, yielding a deposition rate of $\sim$5\,\AA/s. This temperature is well below the bulk boiling point ($\sim$389\,$^\circ$C) and near the melting point ($\sim$212\,$^\circ$C), as per standard pTP datasheets~\cite{ptpDS}. Operating in this regime was found to minimize spitting and produce uniform and optically smooth pTP films (see Fig.~\ref{fig:deposition}).

\paragraph{pTP material and deposition campaigns}

Three deposition campaigns were carried out as the process was optimized. BOROFLOAT\textsuperscript{\textregistered} 33 borosilicate glass (B33) is the baseline optical filter material currently implemented in the DUNE Phase-I photon detection system and therefore serves as the primary substrate of interest in this study. Quartz and sapphire substrates were included to evaluate the general applicability and robustness of the coating technique on alternative optical materials, as well as to explore potential suitability for other detector or photonic applications. The first campaign (Batch 1) used a generic pTP powder on small witness samples of B33, quartz, and sapphire without plasma treatment, and was employed solely to explore basic PVD operating conditions. A second campaign (Batch 2) used full-size 143.75\,mm~$\times$~143.75\,mm substrates (10 sapphire, 15 B33, and 10 quartz) coated to a nominal thickness of 1.5\,\textmu m. For each substrate type, approximately half of the samples received plasma surface treatment and the remainder were processed without it, still using generic pTP. This campaign informed the choice of plasma treatment, thickness range, and chamber conditions for the final recipe.

For the third campaign (Batch 3), the source material was upgraded to high-purity pTP from Thermo Fisher Scientific, and only B33 substrates were coated (15 samples at 1\,\textmu m thickness), all with plasma surface treatment applied. The change from generic to branded high-purity pTP led to noticeably cleaner emission spectra with a more clearly defined peak and higher apparent light yield. Unless otherwise stated, the B33 emission and thickness results presented in the analysis in Sec.~\ref{sec:results} focus on these best-controlled samples: high-purity pTP on B33 from the third campaign, and plasma-treated quartz and sapphire from the second campaign.

\paragraph{Throughput and scaling for mass production}

\begin{figure}[t]
  \centering
  \subfloat[Three-tier rotating dome\label{fig:rotating_dome1}]{
    \includegraphics[width=0.5\textwidth]{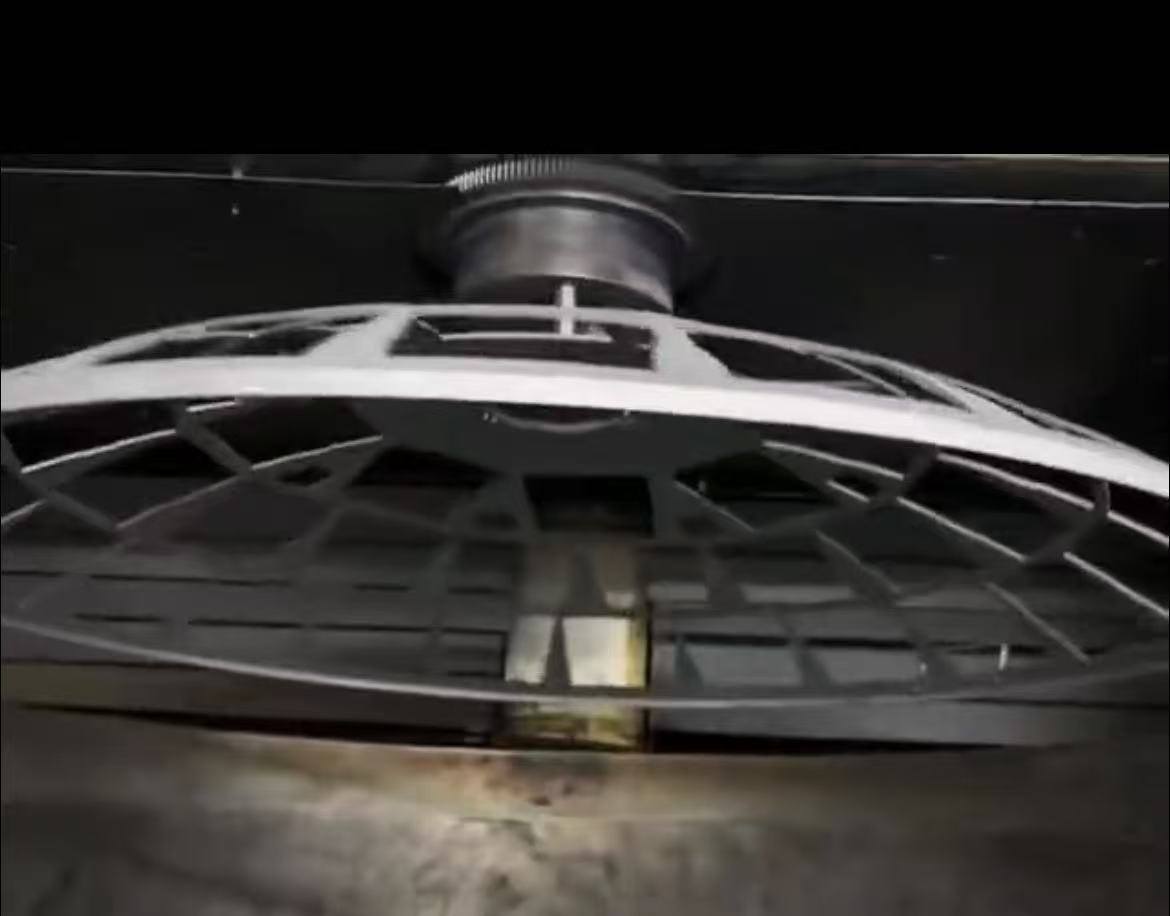}
  }\hfill
  \subfloat[Dome loaded prior to deposition\label{fig:rotating_dome2}]{
    \includegraphics[width=0.35\textwidth]{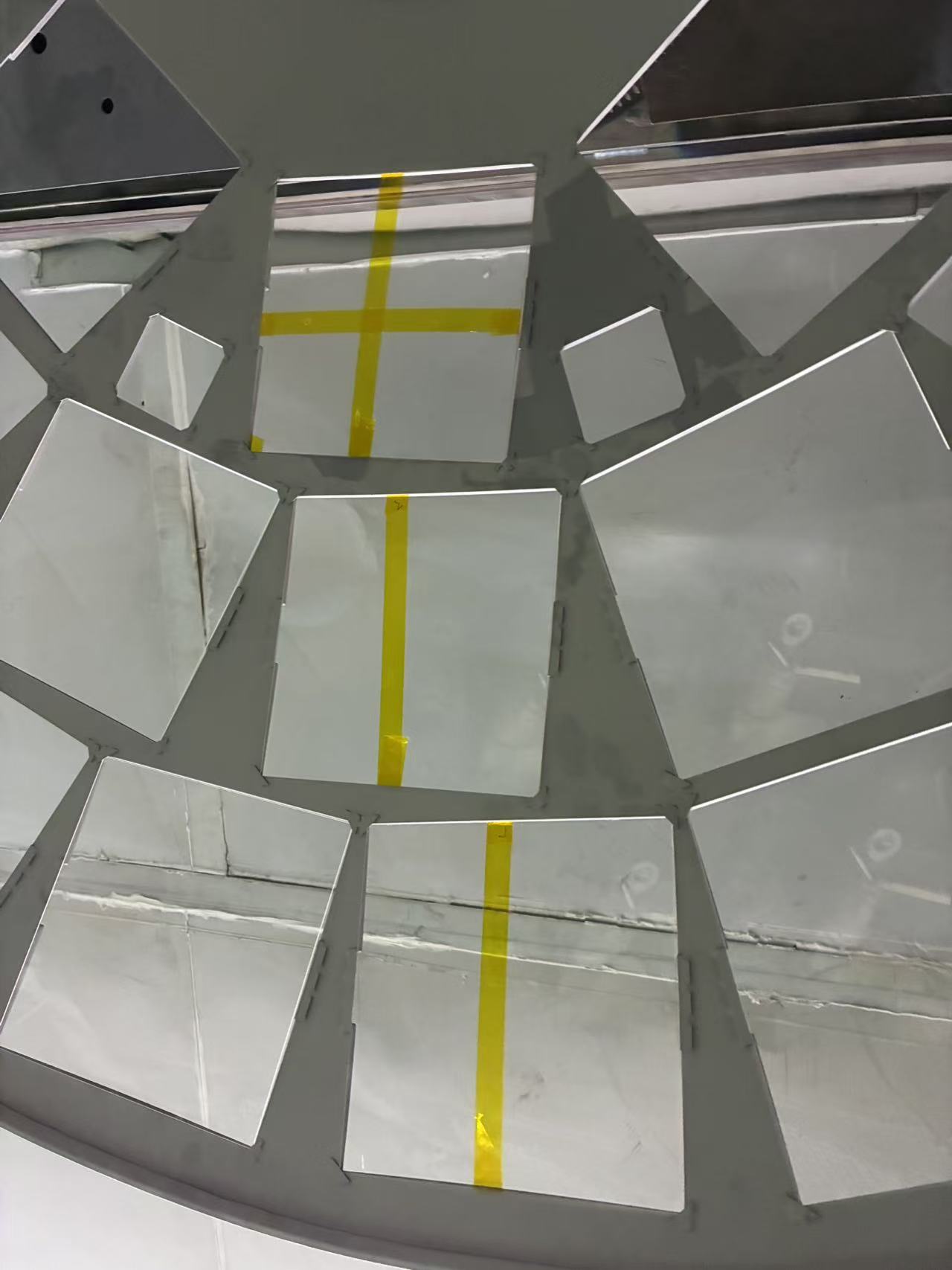}
  }
  \caption{Throughput scaling with a rotating substrate dome. }
  \label{fig:rotating_dome}
\end{figure}

A custom three-tier rotating dome was fabricated for 143.75\,mm\,$\times$\,143.75\,mm substrates (standard DUNE Phase-I PDS size), shown in Fig.~\ref{fig:rotating_dome}. The inner, middle, and outer rings hold 5, 8, and 21 substrates respectively (34 per batch) in a \O\,54\,inch chamber. For a target thickness of 2\,$\mu$m at $\sim$5\,\AA/s, six batches per day under two-shift operation correspond to \(\sim\)5.4\,m$^2$/day. Two coaters with a four-person crew on a dual-shift arrangement would meet an annualized output of $\mathcal{O}$(2000\,m$^2$), satisfying the demand of the DUNE Phase-II Far Detector photon-detection system.

In addition to the deposition step itself, the auxiliary process steps were evaluated for scalability. Substrate cleaning and handling scale linearly with batch size but are not rate-limiting relative to the PVD cycle. Plasma surface treatment is performed with substrates loaded in multiple layers within the plasma chamber, whose capacity exceeds that of the PVD coater; as a result, the nominal 15-minute plasma treatment time remains unchanged when operating at full coater load. Similarly, the in-situ argon ion bombardment step is applied simultaneously to all substrates mounted on the rotating dome and therefore retains a fixed duration of approximately 10 minutes independent of batch size. Consequently, the overall production throughput is dominated by the pTP deposition time, and the additional preparation steps do not introduce a scaling penalty for DUNE Phase-II FD-scale production.

\paragraph{Industrial durability and Quality Control (QC)} Coating durability was evaluated following a subset of the MIL-C-48497A standard for optical coatings\cite{MIL_std}. The pTP films passed adhesion testing (no delamination after rapid removal of applied tape) and moderate abrasion testing (no visible damage after repeated cheesecloth rubbing under pressure). These results establish baseline mechanical robustness consistent with industrial optical-coating practices. 

\subsection{Photo-spectrometer setup and measurement procedure}\label{subsec:spectro}

\begin{figure}[t]
  \centering
  \subfloat[Monochromator\label{fig:monochromator_side}]{
    \includegraphics[width=0.33\textwidth]{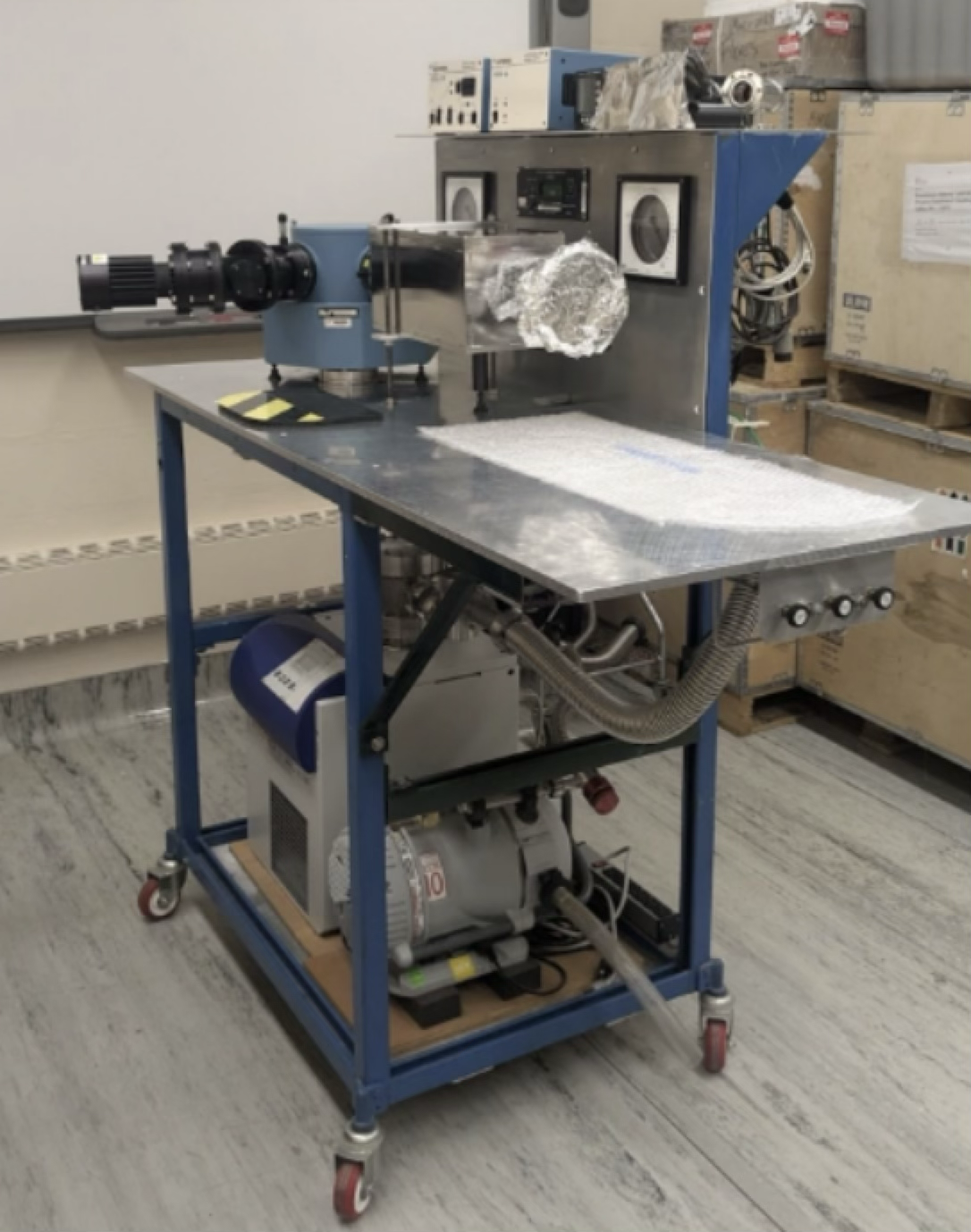}
  }\hfill
  \subfloat[Schematic\label{fig:monochromator_schematic}]{
    \includegraphics[width=0.45\textwidth]{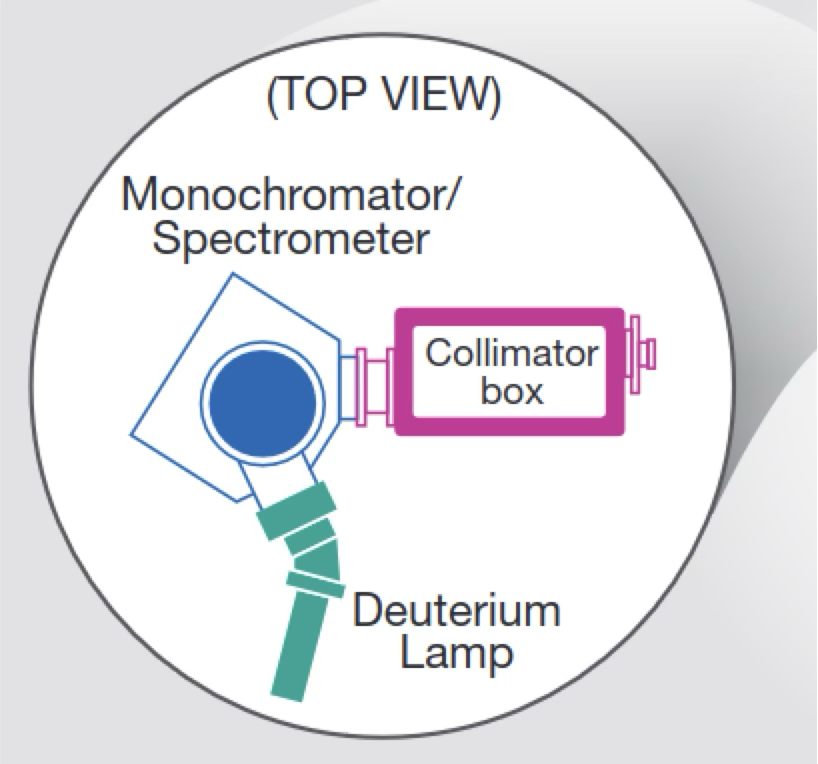}
  }
  \caption{Monochromator setup used for emission-spectrum measurements.}
  \label{fig:monochromator}
\end{figure}

The emission properties of the coatings were characterized using a refurbished McPherson 234/302 VM monochromator/spectrometer equipped with a high-intensity deuterium lamp, providing excitation in the 110--550\,nm range (Fig.~\ref{fig:monochromator}). Initial measurements of fluorescence efficiency and UV transparency were performed at room temperature.%\footnote{Planned cryogenic stress tests will be conducted in liquid nitrogen and in a 260\,L liquid-argon test stand at BNL.}

Emission spectra were recorded with a high-sensitivity cooled-CCD spectrometer (QE Pro 65000 by Ocean Optics). The deuterium lamp output is wavelength-selected by a motorized grating with an effective bandwidth of about 4~nm full width at half maximum (FWHM); the wavelength selection slit is fully opened to maximize excitation light intensity, and the light is then delivered to the sample chamber via a collimated beam path. The optical train operates in an argon-purged enclosure to minimize O$_2$/H$_2$O absorption at short wavelengths.

\subsection{Profilometer measurements}\label{subsec:profilometer}

To measure the thickness of the deposited pTP layer, a stylus profilometer was used. Because the instrument measures relative height differences, each substrate was prepared with a small uncoated region near one edge or corner to provide a distinct step between the bare substrate and the coated surface. This ``step'' was created by masking a narrow strip of the substrate during deposition so that pTP was not deposited in that area. The height difference between the coated and uncoated regions directly corresponds to the film thickness.

After deposition, multiple scans were taken across this masked step to quantify both the average coating thickness and its local variation. Scans typically covered a range of 3.5\,mm from the uncoated edge into the coated region, with the central 0.5--2.5\,mm interval used for uniformity analysis. The profilometer has a vertical resolution of approximately 2\,nm, allowing precise measurement of small thickness variations across the surface. Together, the UV--Vis spectroscopic measurements and profilometer scans provided complementary surface characterization of the coated samples, enabling quantitative assessment of both optical and geometric uniformity. 

\subsection{Tested substrates}\label{subsec:substrates}

As summarized in Sec.~\ref{subsec:procedure}, pTP coatings were deposited on B33 glass, quartz (fused silica), and sapphire substrates as part of the multi-campaign optimization program. Here we briefly summarize the relevant material properties; a comparison is given in Table~\ref{tab:substrate_compare}.

B33 glass is a borosilicate glass widely used for optical and detector applications due to its low thermal expansion, chemical resistance, and relatively low cost. It is commercially available in large, uniform sheets and is straightforward to machine. Transmission in the near-UV is moderate, with increased absorption below $\sim$300\,nm~\cite{schott_b33,schott_b33_latest}. Quartz (fused silica) provides excellent UV transmission to $\sim$160\,nm and very low thermal expansion, supporting repeated cryogenic cycling~\cite{malitson_fusedsilica,Heraeus2023FusedSilica}. The primary limitation is cost for large-area panels. Sapphire (Al$_2$O$_3$) is extremely hard and durable, with broad UV--IR transparency and high thermal conductivity. It presents a stringent test for film adhesion and stability but is relatively expensive and challenging to process in large areas~\cite{Dobrovinskaya2009Sapphire,crystran_sapphire}. Brief summaries and comparative properties are provided in Table~\ref{tab:substrate_compare}.

\begin{table}[h]
\centering
\caption{Comparison of tested substrates. UV cutoff and thermal-expansion values are approximate, compiled from manufacturer datasheets and standard references.}
\resizebox{\textwidth}{!}{%
\begin{tabular}{lcccc}
\toprule
\textbf{Substrate} & \textbf{UV Transparency Cutoff} & \textbf{Thermal Expansion Coefficient} & \textbf{Mechanical Properties} & \textbf{Relative Cost / Availability} \\
\midrule
B33 glass   & $\sim$300\,nm & $\sim$3.2$\times 10^{-6}$/K & Robust, easy to process          & Low, widely available \\
Quartz      & $\sim$160\,nm & $\sim$0.5$\times 10^{-6}$/K & Excellent cryogenic stability    & Medium, limited sheet sizes \\
Sapphire    & $\sim$150\,nm & 5--8$\times 10^{-6}$/K (anisotropic) & Very hard, high durability       & High, expensive, smaller sizes \\
\bottomrule
\end{tabular}%
}
\label{tab:substrate_compare}
\end{table}

\section{Results}\label{sec:results}

Where multiple deposition campaigns exist for the same substrate, we report the results from the best-controlled configuration as summarized in Sec.~\ref{subsec:procedure}. The B33 emission spectrum and thickness measurement results shown here correspond to the optimized third batch, while the quartz and sapphire results are drawn from the plasma-treated subset of the second multi-substrate campaign. 

\subsection{Emission spectrum}\label{subsec:emission}

\begin{figure}[!ht]
  \centering
  \subfloat[Area un-normalized spectra \label{fig:monochromator_spectrum_unnorm}]{
    \includegraphics[width=0.7\textwidth]{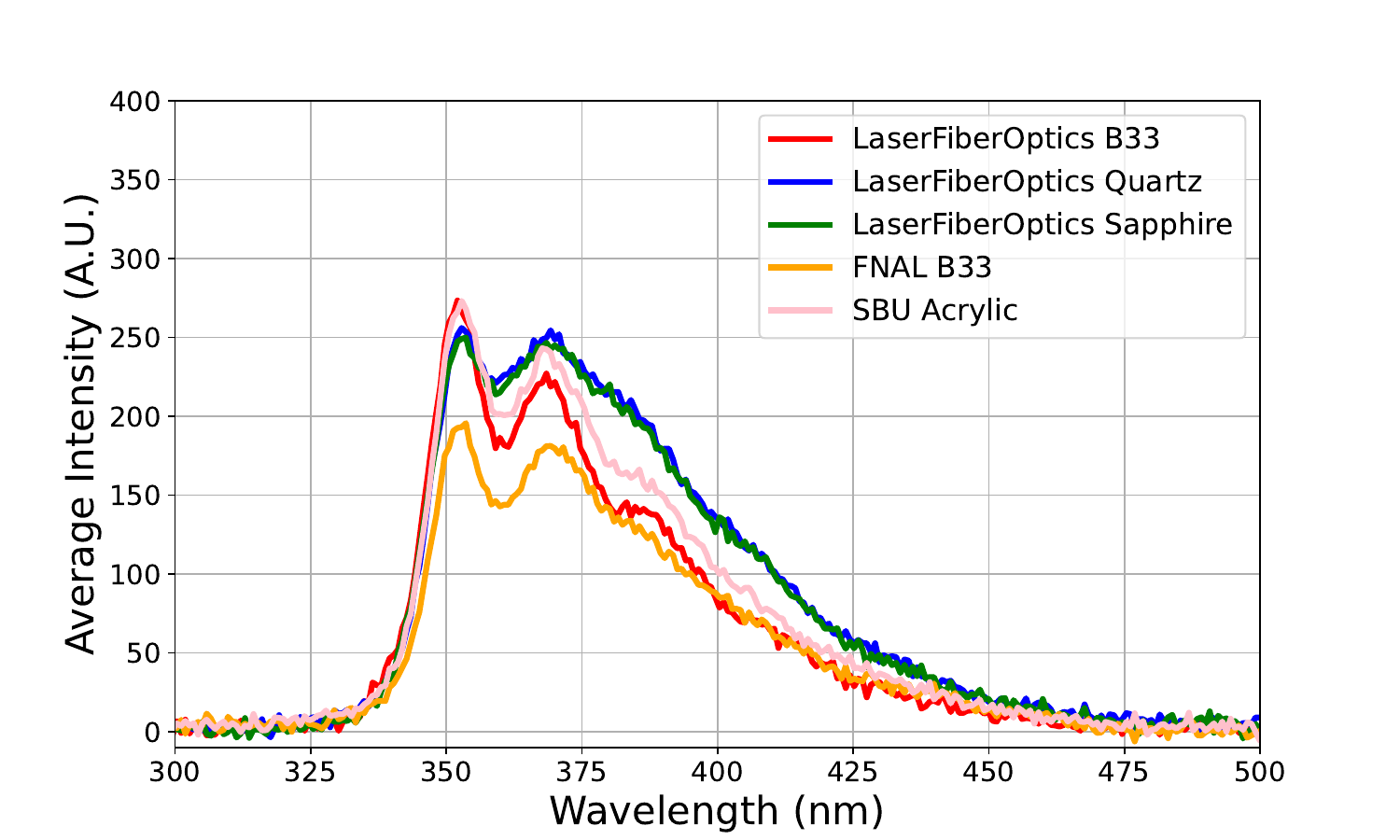}
   }\hfill
  \subfloat[Area normalized spectra \label{fig:monochromator_spectrum_norm}]{
    \includegraphics[width=0.7\textwidth]{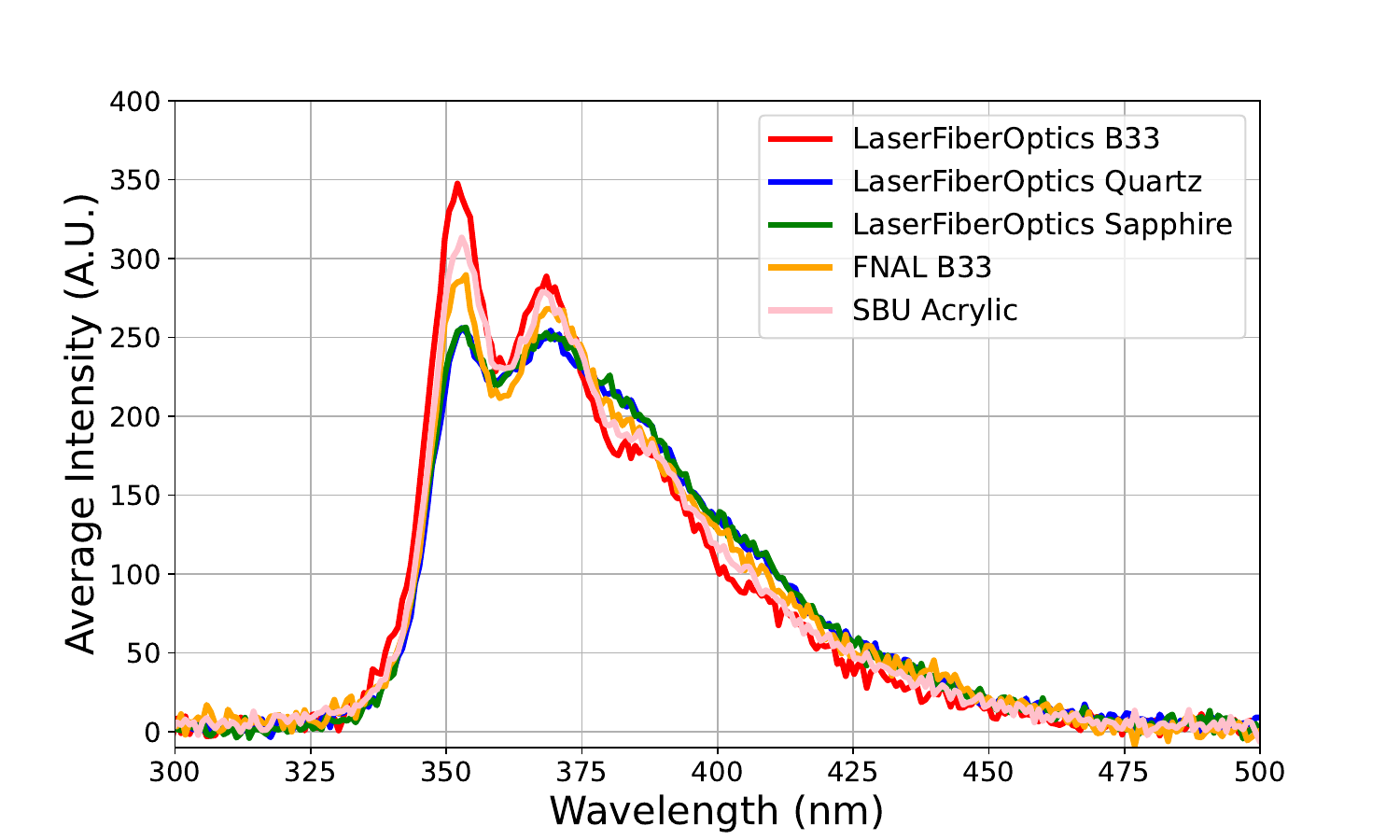}
  }\\
  \caption{Emission spectra measured from pTP-coated samples under UV excitation at 266\,nm. (a) Un-normalized spectra under identical acquisition conditions, enabling relative fluorescence yield comparison. (b) Area-normalized spectra, facilitating spectral shape comparison across samples and substrates.}
  \label{fig:monochromator_spectrum}
\end{figure}

Emission spectra of pTP-coated samples were acquired with long integration times using the monochromator setup described in Sec.~\ref{subsec:spectro}, with monochromatic excitation at 266\,nm selected from a deuterium lamp; a representative set is shown in Fig.~\ref{fig:monochromator_spectrum}. 

pTP has strong absorption in the deep-UV region, peaking near 266\,nm and remaining efficient below 300\,nm as shown in Fig.~\ref{fig:ptp_spectrum}. Therefore, 266\,nm was selected as a common excitation wavelength for emission measurements. For organic wavelength shifters, once the excitation energy exceeds the lowest singlet excited-state threshold, the emission spectrum is independent of excitation wavelength (Kasha's rule \cite{kasha}). As a result, the spectral shape of the excitation source does not contribute to the measured emission spectra, and no unfolding of the lamp spectrum is required. We note that excitation at 127\,nm (LAr scintillation) or 178\,nm (LXe scintillation) may introduce differences in absorption depth and radiative transport. Direct validation under VUV excitation is planned in a future upgrade with the full optical path and sample operated in an argon atmosphere to enable measurements at 127\,nm and 178\,nm.

The measured emission profiles are in agreement with the canonical pTP spectrum (Fig.~\ref{fig:ptp_spectrum}), both in peak position (approximately 350\,nm) and overall spectral shape. Spectra obtained from independent reference samples produced at Fermilab and Stony Brook University (SBU) exhibit the same features. The Fermilab reference sample consists of Thermo Fisher pTP deposited by PVD onto a B33 glass substrate following the DUNE Phase-I design (including a dichroic filter), while the SBU reference sample uses Thermo Fisher pTP deposited by a solvent-based spray process onto an acrylic substrate. The close agreement among these spectra, despite differences in substrate and deposition technique, confirms that the industrially deposited films preserve the intrinsic optical behavior of pTP.

Since all samples were measured under identical acquisition conditions, the un-normalized spectra in Fig.~\ref{fig:monochromator_spectrum_unnorm} provide a relative fluorescence yield comparison. The B33 samples from the third deposition campaign (high-purity Thermo Fisher pTP) show fluorescence yields broadly comparable to the reference samples, and slightly higher than the FNAL B33 reference; the latter difference may be partly attributable to the dichroic filter present in the FNAL sample. The quartz and sapphire samples from the second campaign show moderately different spectral shapes compared to the B33 and reference samples, consistent with the use of generic pTP source material in that campaign, as discussed below.

Peak variation among the B33 samples from the Batch 3 and reference samples was less than 5\,nm, indicating strong reproducibility across substrates and deposition batches. This consistency demonstrates that neither substrate composition nor plasma pre-treatment measurably alters the photophysical response of high-purity pTP under the conditions studied here.

During earlier process-development runs using a generic pTP powder (see Fig.~\ref{fig:monochromator_spectrum_norm}), we observed less well-defined emission features and lower apparent light yield, indicating that control of the pTP source purity is an important factor in achieving reliable, high-quality, and reproducible emission spectra. This observation is consistent with the purity requirements discussed in Sec.~\ref{sec:intro}.

\subsection{Thermal and cryogenic stress response}\label{subsec:thermal}

To assess the robustness of the wavelength-shifting films under realistic detector conditions, preliminary thermal and cryogenic stress tests were performed using B33 samples from the second deposition campaign (see Sec.~\ref{subsec:procedure}). These samples, which employed a generic pTP source and included plasma-treated substrates, were used for initial robustness studies so that the final high-purity B33 batch could be reserved for detailed optical characterization.

Three samples were selected for cryogenic test by immersion into liquid nitrogen (LN$_2$) using an open-mouth dewar used for DUNE cold electronics testing. Both vertical and horizontal orientation measurements were conducted to simulate filter configurations in large detectors. For the vertical configuration, samples were placed in a basket above the LN$_2$ surface and pre-cooled for approximately 30 minutes. The basket was then slowly lowered into the liquid over $\approx$ 10 minutes to minimize thermal shock. The samples remained fully immersed for $\approx$ 18 hours. After immersion, they were raised in stages above the liquid surface to allow gradual warming, removed once the surface temperature exceeded 0~$^{\circ}$C, dried with forced air, and subsequently placed in a $\sim$30~$^{\circ}$C dryer for 1 hour. For the horizontal configuration, the same immersion and warming procedure was followed with a longer pre-cooling time of $\approx$ 1 hour prior to immersion to further reduce thermal stress in the horizontal arrangement. This test was conducted after the substrates had been warmed up for 24 hours in air after completely drying.

\begin{figure}[t]
  \centering
  \includegraphics[width=0.85\textwidth]{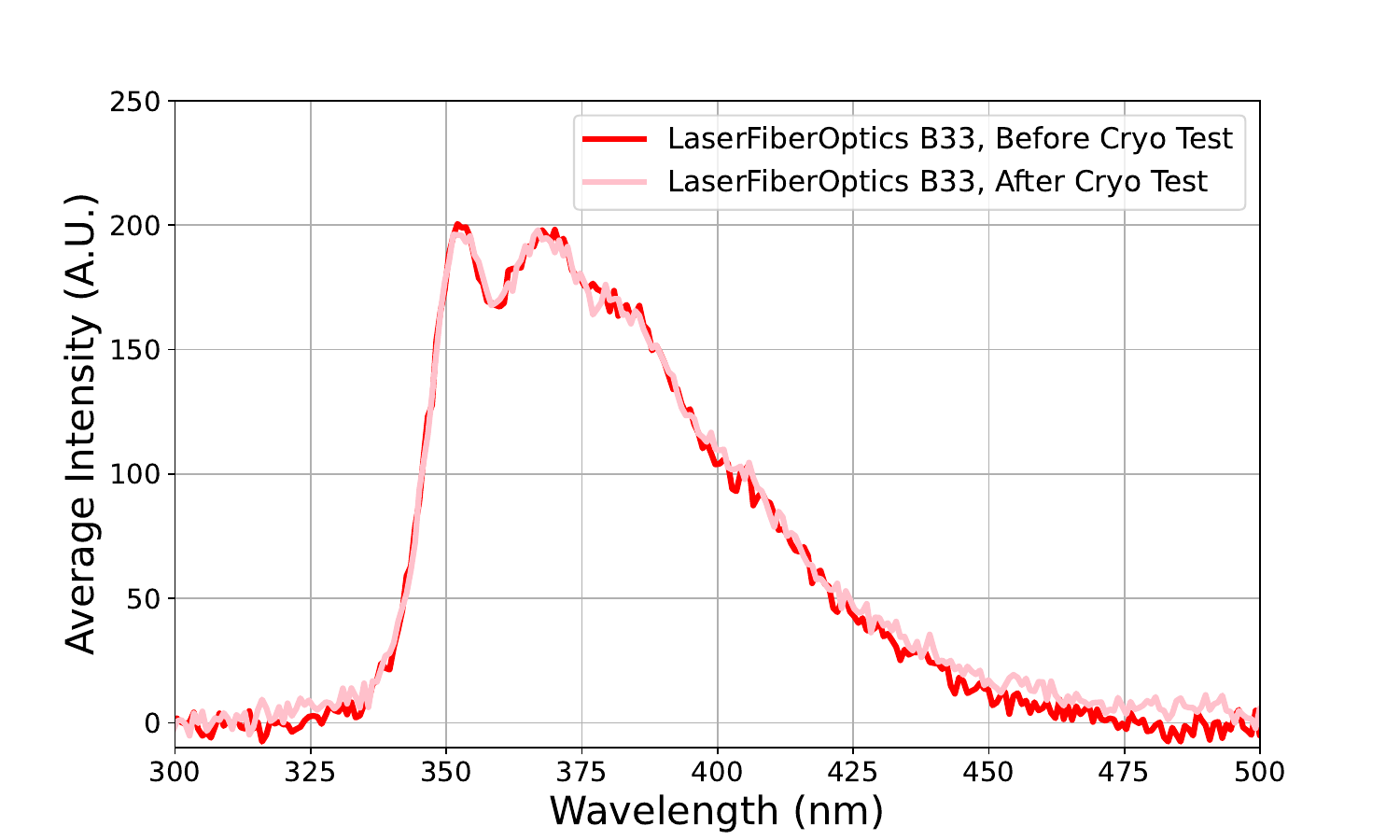}
  \caption{Comparison of emission spectra before and after the cryogenic cycle.}
  \label{fig:cryocycle}
\end{figure}

After multiple cryogenic cycles, no macroscopic delamination, peeling, or cracking was observed. Minor visual changes were noted on quartz and sapphire substrates during warm-up, possibly associated with surface condensation or transient moisture effects rather than degradation of the coating itself. A direct comparison of the emission spectra measured before and after thermal cycling under identical, non-normalized acquisition settings for one B33 sample is shown in Fig.~\ref{fig:cryocycle}. Within experimental uncertainty, no significant change in spectral shape or emission intensity was observed. While these results are encouraging, we emphasize that this study represents a preliminary assessment based on a limited number of samples. More controlled long-duration cryogenic tests, including operation in a moisture-free environment, additional profilometric surface characterization, and studies across a larger number of tiles, are planned to further evaluate long-term robustness for large-scale LArTPC deployment.

\subsection{Coating thickness and uniformity}\label{subsec:thickness}

\begin{figure}[t]
  \centering
  \subfloat[Thickness vs.\ lateral distance (B33)\label{fig:height_v_lateral_B33}]{
    \includegraphics[width=0.48\textwidth]{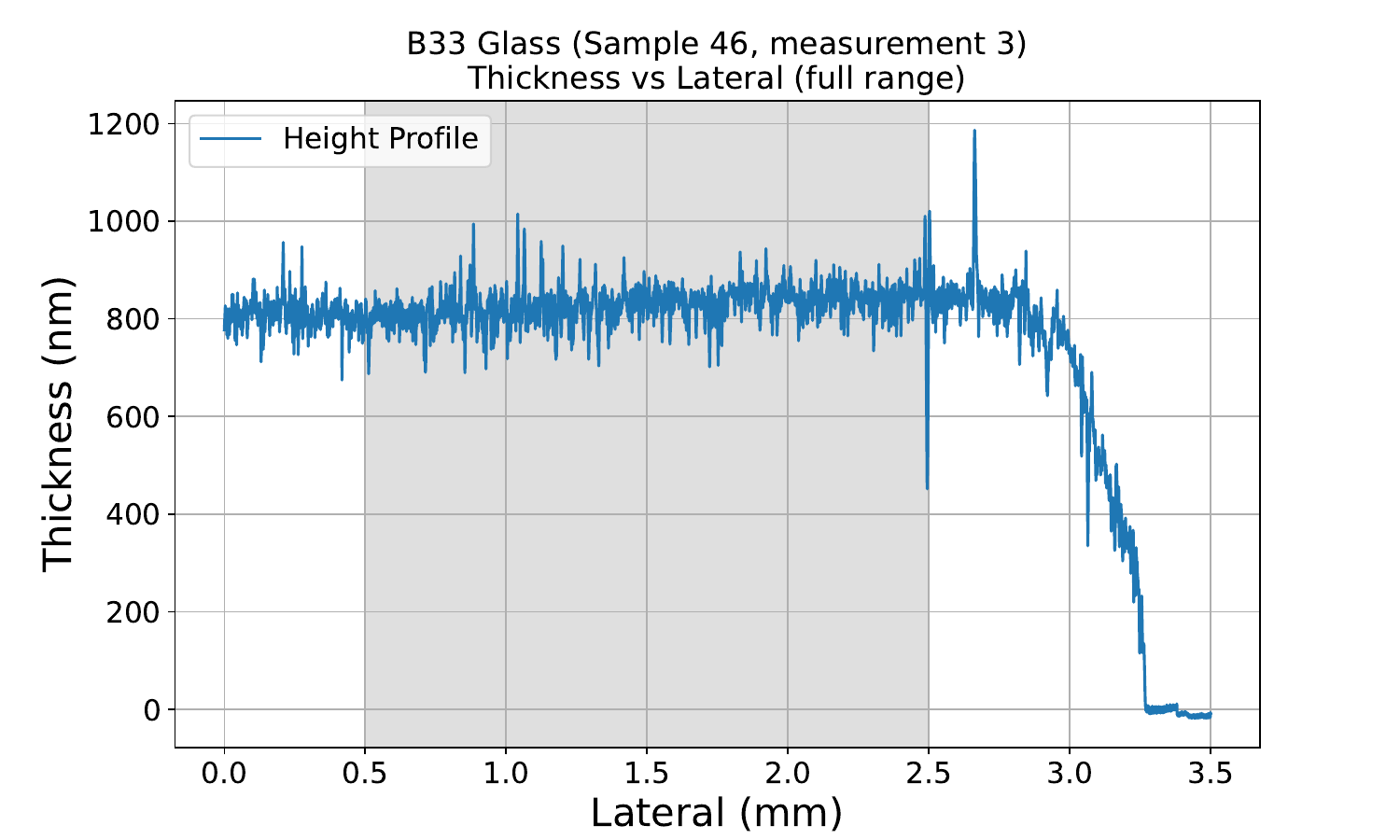}
   }\hfill
  \subfloat[Thickness distribution (B33)\label{fig:thickness_dist_B33}]{
    \includegraphics[width=0.48\textwidth]{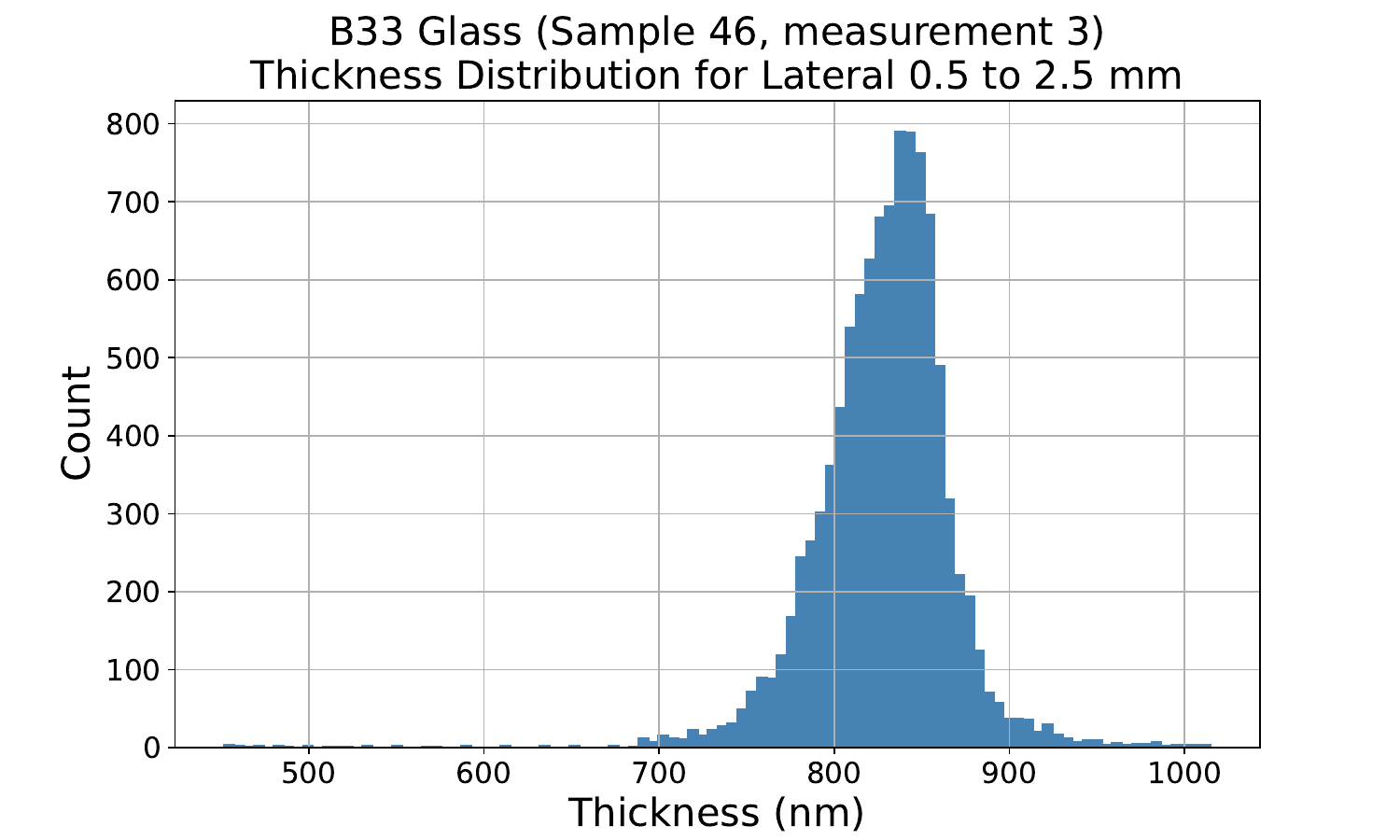}
  }\\
  \subfloat[Thickness vs.\ lateral distance (Sapphire)\label{fig:height_v_lateral_sapphire}]{
    \includegraphics[width=0.48\textwidth]{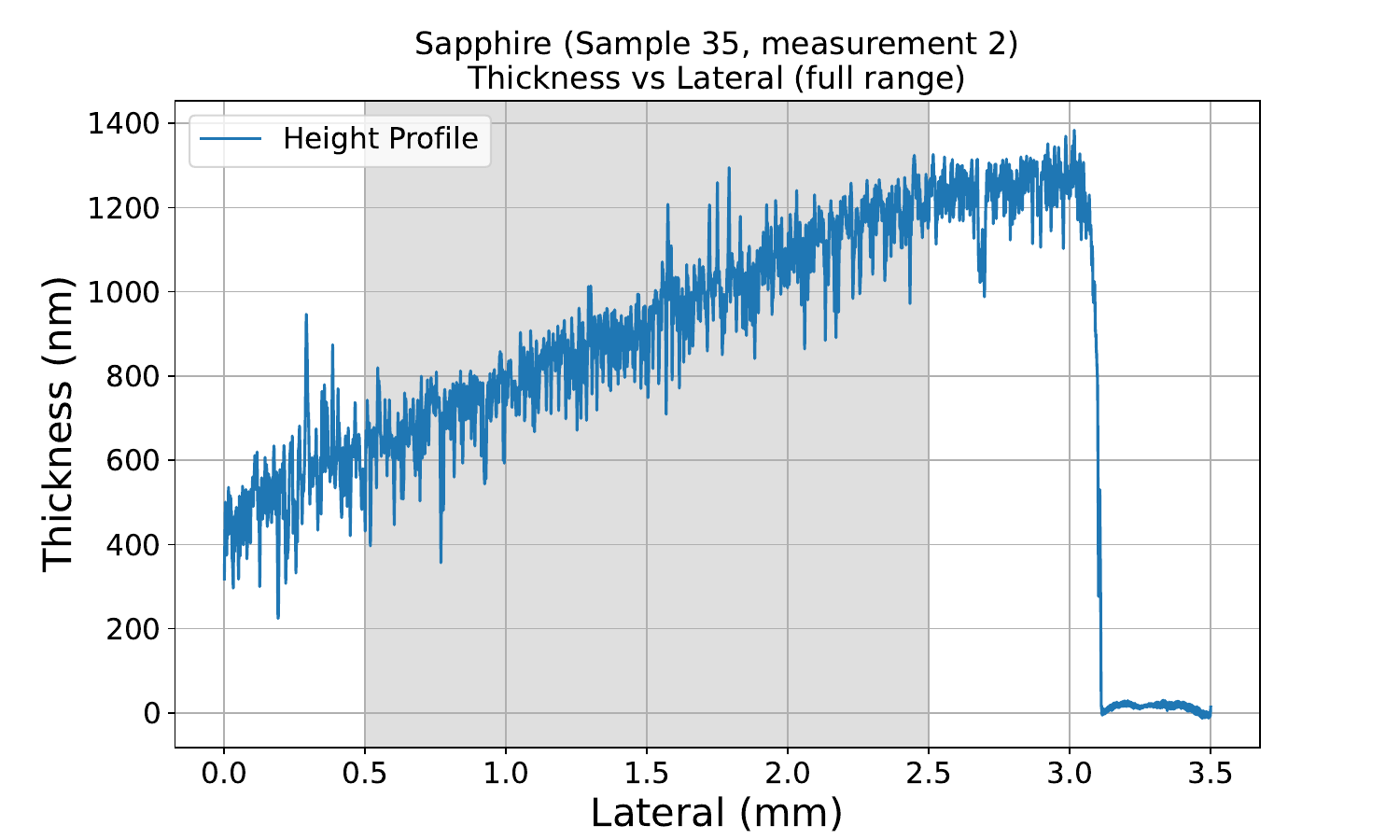}
  }\hfill
  \subfloat[Thickness distribution (Sapphire)\label{fig:thickness_dist_sapphire}]{
    \includegraphics[width=0.48\textwidth]{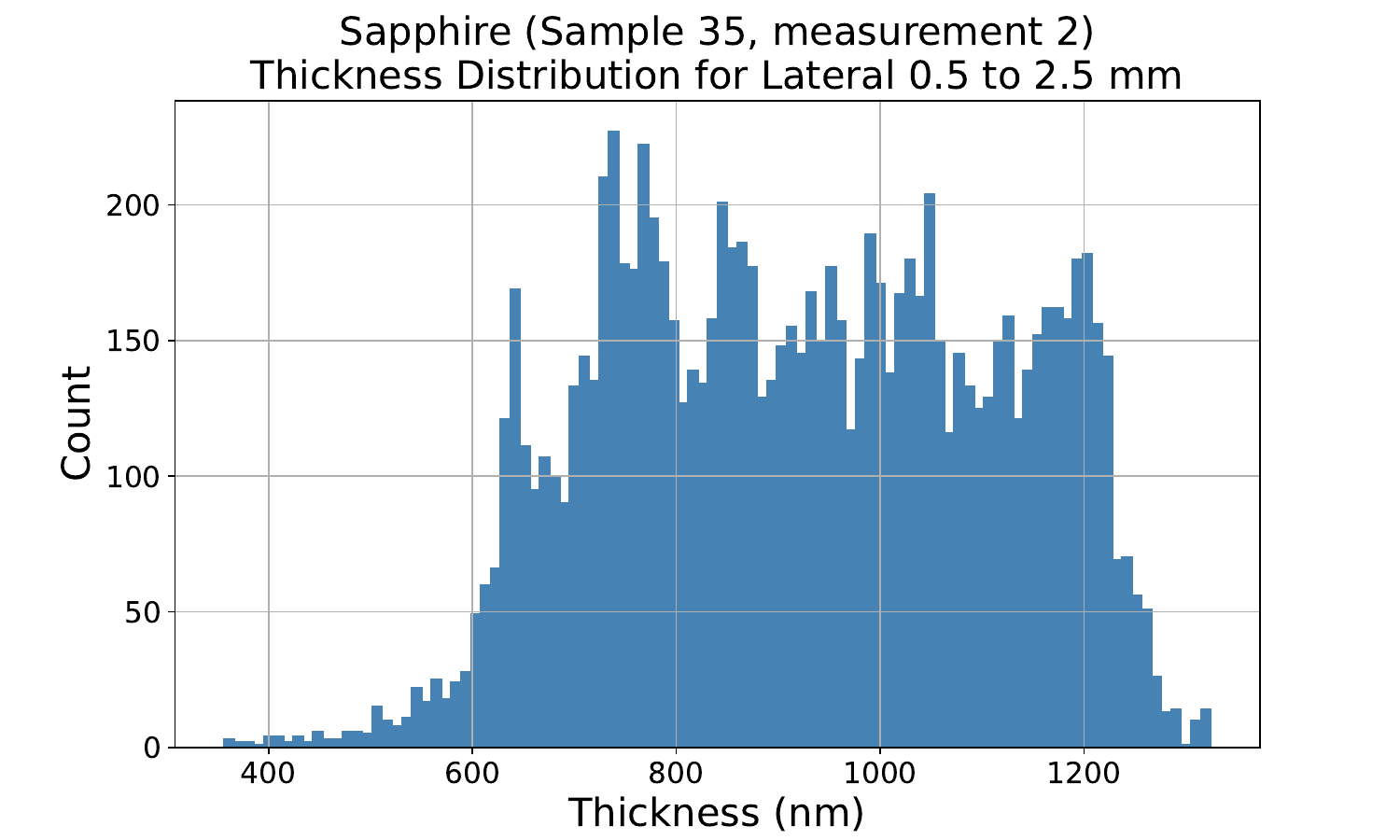}
  }\\
  \subfloat[Thickness vs.\ lateral distance (Quartz)\label{fig:height_v_lateral_quartz}]{
    \includegraphics[width=0.48\textwidth]{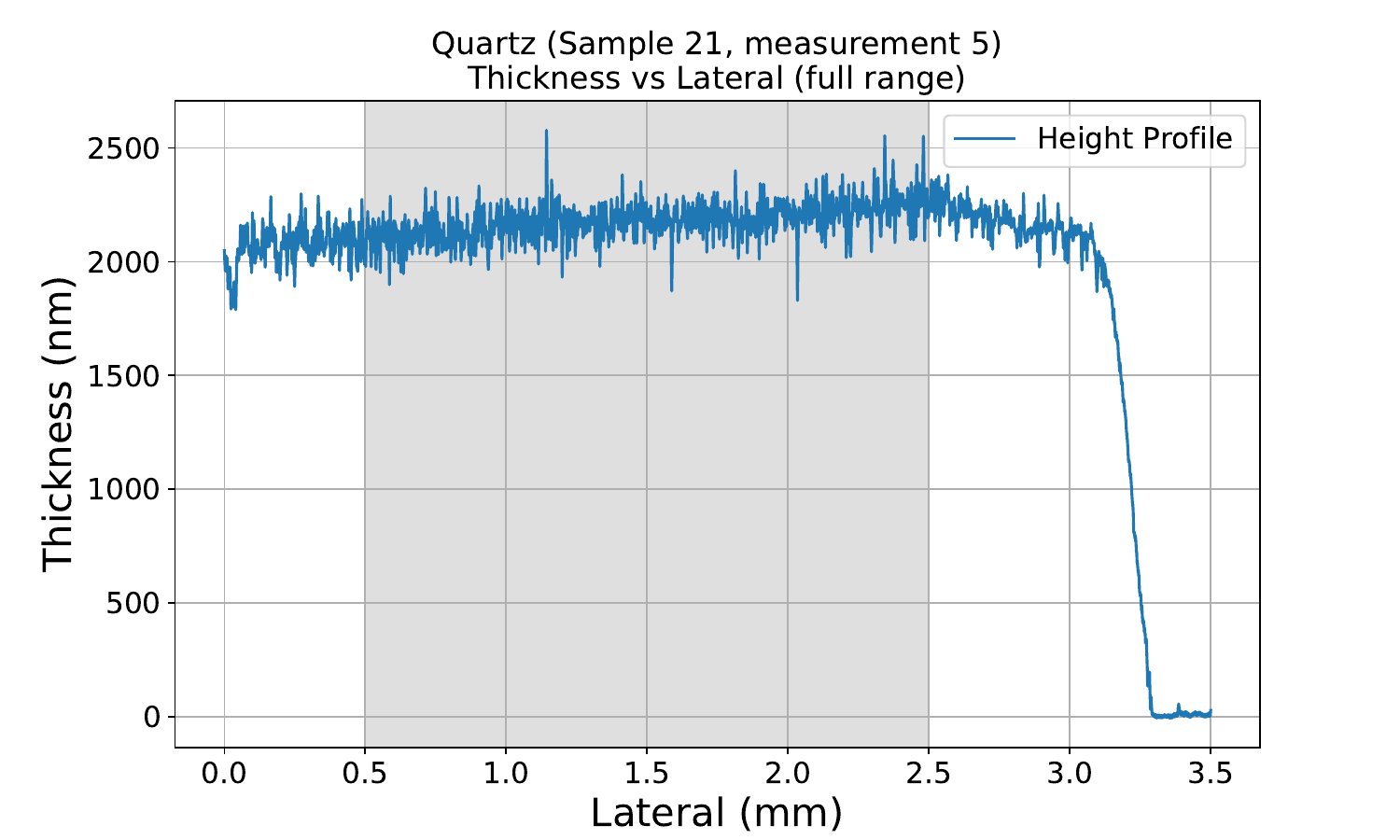}
  }\hfill
  \subfloat[Thickness distribution (Quartz)\label{fig:thickness_dist_quartz}]{
    \includegraphics[width=0.48\textwidth]{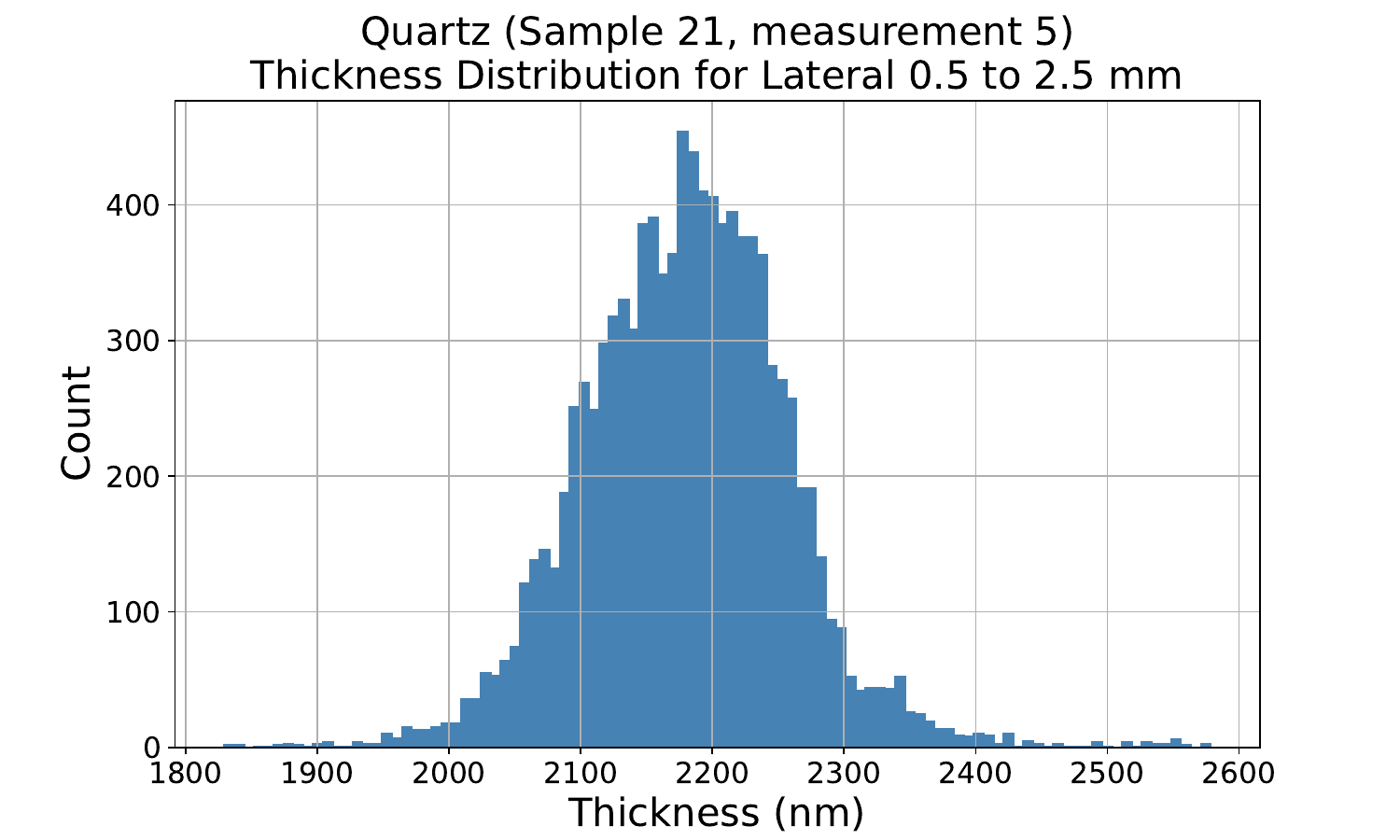}
  }
  \caption{Representative profilometer scans for each substrate type. Left: thickness vs.\ lateral scan distance, where the gray band indicates the lateral range used for thickness statistics (0.5--2.5\,mm from the step edge). Right: thickness distribution over the same interval. (a, b) B33 glass (Sample S46, Batch 3); (c, d) Sapphire (Sample S35, Batch 2); (e, f) Quartz (Sample S21, Batch 2). The apparent slope in the sapphire scan (c) is consistent with substrate bow rather than coating non-uniformity. Full thickness statistics across all measured samples are reported in Table~\ref{tab:thickness}.}
  \label{fig:profilometer_result}
\end{figure}

Film thickness was measured using the stylus profilometer described in Sec.~\ref{subsec:profilometer}, by scanning across a masked step between the uncoated and coated regions of each substrate; the central 0.5--2.5,mm interval of each scan was used for thickness statistics. For Batch 2 samples, six scans at different edge positions distributed around the substrate perimeter were taken; for Batch 3 samples, up to four corner positions were available due to the placement of the masked step. Table~\ref{tab:thickness} reports statistics computed across all valid scans for every measured sample in both campaigns; samples excluded from the table and the reasons for exclusion are noted in the caption. Two complementary statistics are reported per sample: the scan-to-scan standard deviation, quantifying thickness variation between different edge positions on the same substrate, and the per-scan standard deviation (reported as a min--max range across scans), quantifying local height fluctuation within each individual 0.5--2.5\,mm scan interval. We note that the stylus profilometer is limited in travel range to mm-scale scans near the masked step edge, and therefore these measurements do not constitute a full-area uniformity map across the 143.75\,mm$\times$143.75\,mm substrate.

\begin{table}[t]
    \centering
    \caption{Film thickness statistics per sample from profilometry measurements. For each sample, the mean thickness and scan-to-scan standard deviation are computed across all valid edge scans. The per-scan std dev range (min--max) reflects local thickness variation within each 0.5--2.5\,mm scan interval. Batch 2 samples were measured at 6 edge positions; Batch 3 samples were measured at up to 4 corner positions. Samples S40 and S45 have fewer scans due to insufficiently large masked steps at some corners. Samples S41, S47, S48, S49, and S50 were excluded due to suspected instrumental issues during measurement; S56 was excluded as it was deposited at a different target thickness of $\sim$4\,$\mu$m.}
    \label{tab:thickness}
    \resizebox{\textwidth}{!}{
    \begin{tabular}{llllccc}
        \hline
        Batch & Sample & Substrate & No.\ scans & Mean thickness (nm) & 
        Scan-to-scan std dev (nm) & Per-scan std dev range (nm) \\
        \hline
        2 & S9  & B33      & 6           & 1758.5 & 210.9 & 50.8\,--\,396.6 \\
        2 & S12 & B33      & 6           & 1585.6 &  46.6 & 56.3\,--\,108.4 \\
        2 & S15 & B33      & 6           & 2115.3 & 165.2 & 41.6\,--\,239.4 \\
        2 & S20 & B33      & 6           & 1526.2 & 136.9 & 54.4\,--\,147.8 \\
        \hline
        3 & S40 & B33      & 3$^\dagger$ &  987.1 &  72.6 & 56.5\,--\,173.3 \\
        3 & S42 & B33      & 4           &  990.5 &  18.5 & 50.9\,--\,131.8 \\
        3 & S43 & B33      & 4           & 1001.3 &  39.4 & 27.5\,--\,173.9 \\
        3 & S44 & B33      & 4           &  960.3 &  47.7 & 61.4\,--\,140.1 \\
        3 & S45 & B33      & 2$^\dagger$ &  965.9 &  36.4 & 41.5\,--\,55.4  \\
        3 & S46 & B33      & 4           &  955.1 &  88.7 & 33.8\,--\,137.2 \\
        \hline
        2 & S21 & Quartz   & 6           & 2124.0 & 160.0 & 77.9\,--\,297.0 \\
        2 & S29 & Quartz   & 6           & 1336.1 &  89.1 & 65.6\,--\,229.1 \\
        \hline
        2 & S2  & Sapphire & 6           & 1562.3 & 187.1 & 118.1\,--\,491.8 \\
        2 & S35 & Sapphire & 6           & 1028.6 & 300.5 & 119.8\,--\,514.2 \\
        \hline
    \end{tabular}
    }
    \begin{flushleft}
        $^\dagger$ Fewer scans available due to insufficiently large masked step at some corners.
    \end{flushleft}
\end{table}

Figure~\ref{fig:profilometer_result} illustrates typical scans. For sapphire, an apparent slope in the thickness vs.\ lateral distance is consistent with substrate bow rather than coating non-uniformity, complicating direct cross-sample comparisons. This is also reflected in the large per-scan standard deviation range observed for sapphire samples in Table~\ref{tab:thickness}, where individual scans at different edge positions differ substantially due to substrate bow rather than actual coating non-uniformity. The full set of per-sample statistics across all valid scans is reported in Table~\ref{tab:thickness}.

All coatings reached the target thickness range of 1--2\,\textmu m. Batch 3 B33 samples show highly consistent mean thicknesses in the range 955--1001\,nm across six samples, with scan-to-scan standard deviations of 18--89\,nm, demonstrating good reproducibility across the batch. Batch 2 B33 samples span a broader range of 1526--2115\,nm, reflecting the wider nominal thickness target of 1.5\,\textmu m used in that campaign. Quartz samples range from 1336--2124\,nm, and sapphire from 1029--1562\,nm. The per-scan standard deviation for B33 samples is generally well below 10\% of the mean thickness across the scanned edge regions, while sapphire exhibits larger spread attributable primarily to substrate bow rather than coating non-uniformity. We note, however, that these uniformity statistics are derived from mm-scale scans near the substrate edges and do not characterize the full 143.75\,mm $\times$ 143.75\,mm substrate area.

Such consistency across different substrates and between batches is encouraging from a manufacturing standpoint, though full-area uniformity mapping across the bulk of each substrate remains to be demonstrated. 

\section{Conclusions}\label{sec:conclusion}

This work represents, to our knowledge, the first successful demonstration of industrial-scale physical vapor deposition of $p$-terphenyl coatings on multiple optical substrates, including borosilicate glass, quartz, and sapphire, with comprehensive surface characterization confirming that the films are smooth, uniform, and adherent, and meet the optical and geometric quality requirements demonstrated in this manufacturing and process study. The industrial process yields consistent 1--2\,\textmu m coatings with sub-10\% uniformity and has proven cost-effective and highly scalable; the demonstrated throughput indicates that the full 2000\,m$^2$ coating required for DUNE could be completed within approximately one year of production.

Beyond demonstrating large-scale manufacturability, this work addresses a long-standing materials challenge: achieving uniform, adherent organic coatings on inorganic substrates. The success of the industrial deposition process shows that interface control through optimized plasma and ion treatments can effectively overcome adhesion and uniformity issues that have traditionally limited such systems. This capability validates the robustness of the present approach and opens the path to broader applications of organic scintillators and wavelength-shifting films in cryogenic detector environments. Ongoing work focuses on further improving surface uniformity to enable next-generation light-trapping geometries with enhanced photon-collection efficiency. Future work will also include quantitative measurements of VUV photon conversion efficiency at 127\,nm and\,178 nm (relevant to Xe-doped LAr scenarios), which will connect the demonstrated film quality to detector-level light-yield requirements.

\acknowledgments
This work is supported by the Laboratory Directed Research and Development (LDRD) program at Brookhaven National Laboratory and by the U.S.\ Department of Energy, Office of Science, Office of High Energy Physics. The authors would also like to thank our industrial partner, LaserFiberOptics LLC for their dedicated collaboration and technical contributions.

\bibliographystyle{unsrt}
\bibliography{TS_JINST}{}

\end{document}